      [1999/12/01 v1.4c Il Nuovo Cimento]
\def\ltsima{$\; \buildrel < \over \sim \;$}
\def\gtsima{$\; \buildrel > \over \sim \;$}
\def\simlt{\lower.5ex\hbox{\ltsima}}
\def\simgt{\lower.5ex\hbox{\gtsima}}
\def\pmb#1{\setbox0=\hbox{#1}%
    \kern-.025em\copy0\kern-\wd0
    \kern.05em\copy0\kern-\wd0
    \kern-.025em\raise.0433em\box0 }

\documentclass{cimento}

\usepackage{graphicx}  
\title{The Future of Cosmology}
\author{George~Efstathiou}
\instlist{\inst{ }Institute of Astronomy, Madingley Road, Cambridge,
CB3 OHA. England.}
\begin{document}

\maketitle

\begin{abstract}
  This article is the written version of the closing talk presented at
  the conference `{\it A Century of Cosmology}' held at San Servolo,
  Italy, in August 2007.  I focus on the prospects of constraining
  fundamental physics from cosmological observations, using the search
  for gravitational waves from inflation and constraints on the
  equation of state of dark energy as topical examples. I argue that
  it is important to strike a balance between the importance of a
  scientific discovery against the likelihood of making the discovery
  in the first place. Astronomers should be wary of embarking on large
  observational projects with narrow and speculative 
  scientific
  goals. We should maintain a diverse range of research programmes as
  we move into a second century of cosmology. If we do so, discoveries
  that will reshape fundamental physics will surely come.

\end{abstract}

\section{Introduction}
It is a privilege to be invited to give the closing talk at this
meeting celebrating {`\it A Century of Cosmology'}. I have taken the
liberty of changing the title of the written version to make it shorter and
snappier. Of course, I will not be able to cover all of cosmology
and I am not a clairvoyant. What I will try to do is to review a small
number of topics and use them as a guide  to how our subject might
develop. I am told that a good high court judge leaves
everybody in the courtroom dissatisfied. I have borne this in mind in
preparing this talk. Rather than congratulating our community on its
remarkable progress, I have tried to pose some difficult questions.
If readers are dissatisfied with some, but not all, of my answers, I
will have succeeded with this talk.

It is a cliche, but nonetheless true, that our subject has undergone a
revolution in the last decade. When I was a graduate student,
astronomers were still struggling with photographic plates and had
measured redshifts for only a thousand or so galaxies. A galaxy
redshift of a half was regarded as exotically high -- almost
primordial. Extragalactic astronomy from space was in its infancy -- a
few bright X-ray sources and little more. The contrast with the
present day is striking. We now have space satellites covering radio
frequencies through to gamma rays. Observational cosmology has moved
with a vengeance into survey mode, with large teams dedicated to
special projects such as surveys of galaxy redshifts, distant
supernovae, weak lensing, and so on. The era of projects geared
towards `precision cosmology' is upon us, with WMAP providing the
archetypal example. Many of the cosmological parameters that we have
been struggling to measure for decades are now constrained accurately,
possibly to higher precision than many astronomers actually care
about. The talks at this meeting provide ample testament to how
cosmology has changed.

Cosmology has clearly been successful and we have learned a lot about
our weird and wonderful Universe. But has hubris set in? Have our
successes led to unachievable expectations concerning the future?
(`Irrational exuberance' as Alan Greenspan famously referred to the
dot-com boom). I will try to answer these questions by considering
three topics in this article. The first two, the search for
gravitational waves from inflation and constraints on the equation of
state of dark energy, use cosmology to test fundamental physics.  The
third topic, the non-linear Universe, may or may not lead to new
results of relevance to fundamental physics. If not, does this mean
that a topic such as galaxy formation is a less worthy problem than
understanding dark energy or inflation -- a cosmological equivalent of
weather forecasting? To what extent should we judge projects by their
potential to test fundamental physics rather than to improve our
understanding of complex non-linear phenomena? These are difficult,
and for some people emotive, questions. But we must face up to them as
we move into a second century of cosmology.

\section{The Search for Tensor Modes}

Inflation is a compelling theoretical idea. The almost perfect
agreement between the cosmic microwave background (CMB) anisotropy
measurements (particularly by WMAP) and theoretical predictions based
on inflation has been interpreted by many, though not all,
cosmologists as evidence that inflation actually happened. A key
prediction of inflationary models is the existence of a stochastic
background of gravitational waves. Such a background is potentially
detectable via a `$B$-mode' polarization signature in the
CMB~\cite{ref:zald, ref:kam}. If only we could measure this $B$-mode
signature, so the thinking goes, we would have incontrovertible
evidence for inflation that should convince the staunchest of
skeptics. A detection of tensor modes would set strong constraints on
the dynamics of inflationary models and would fix the energy scale of
inflation via,
\begin{equation}
V^{1/4} \approx 3.3 \times 10^{16} r^{1/4} \; {\rm GeV}, \label{theory1}
\end{equation}
where $r$ is the tensor-scalar ratio (defined so that for $r \approx
1$ tensor and scalar modes contribute nearly equal amplitudes to the
large angle temperature anisotropies, see
\cite{ref:peiris} for a precise definition).

However, the problem is not a simple one for at least three reasons:

\smallskip

\noindent
{\bf [A] The expected signal will be small:}

\smallskip

Firstly, any $B$-mode signal from inflation will be incredibly small
and difficult to detect.  The direct upper limit on $B$-modes from
WMAP polarization measurements corresponds to a $1\sigma$ upper limit
of about $r \simlt 1$ \cite{ref:page}. From parameter fitting it is
possible to set a $2\sigma$ indirect limit of $r \simlt 0.36$ with
plausible theoretical assumptions \cite{ref:seljak}. A primordial
$B$-mode of this amplitude would produce an {\it rms} anisotropy
signal of only $\sim 0.35 \mu K$, {\it i.e.} about a factor of 20
times smaller than the {\it rms} anisotropy in $E$-modes. This is well
below the sensitivity levels achievable by WMAP. The Planck satellite
\cite{ref:bluebook} scheduled for launch in $2008$ will be
considerably more sensitive than WMAP but will struggle to constrain
$B$-mode anisotropies. 

This is illustrated in Figure 1 which shows power spectra from a
simulation matched to the sensitivity level of the Planck $143$ GHz
polarized detectors. (This simulation has been done at half Planck
resolution, but with the noise level adjusted to match the sensitivity
expected for the $143$ GHz detectors). The simulation employs a
realistic focal plane geometry, scan strategy, low frequency detector
noise and Galactic mask, as described in \cite{ref:gpe1}. The
$B$-mode spectrum for $r=0.15$ becomes noise dominated at $\ell \simgt
20$. Even if all sources of systematic error are kept under control,
Planck will struggle to detect anything more than a few multipoles in
the $B$-mode even if the tensor amplitude is as high as $\sim 0.15$.

Improving the sensitivity further requires large {\it arrays} of detectors.
Several ground based/sub-orbital $B$-mode polarization experiments are
either planned or in progress (examples include BICEP \cite{ref:yoon},
QUIET \cite{ref:seiffert}, SPIDER \cite{ref:montroy} and CLOVER
\cite{ref:taylor}.) Groups in both the USA and Europe have considered
designs for a $B$-mode optimised low resolution ($\simgt 30^\prime$)
space satellite. It is certainly possible to conceive of experiments with
the raw sensitivity to probe tensor-scalar ratios of $r \simlt
10^{-2}$. However, it is not yet clear whether systematic errors can
be reduced to below this level.

\begin{figure}
\centering
\includegraphics[width=0.6\textwidth, angle=-90]{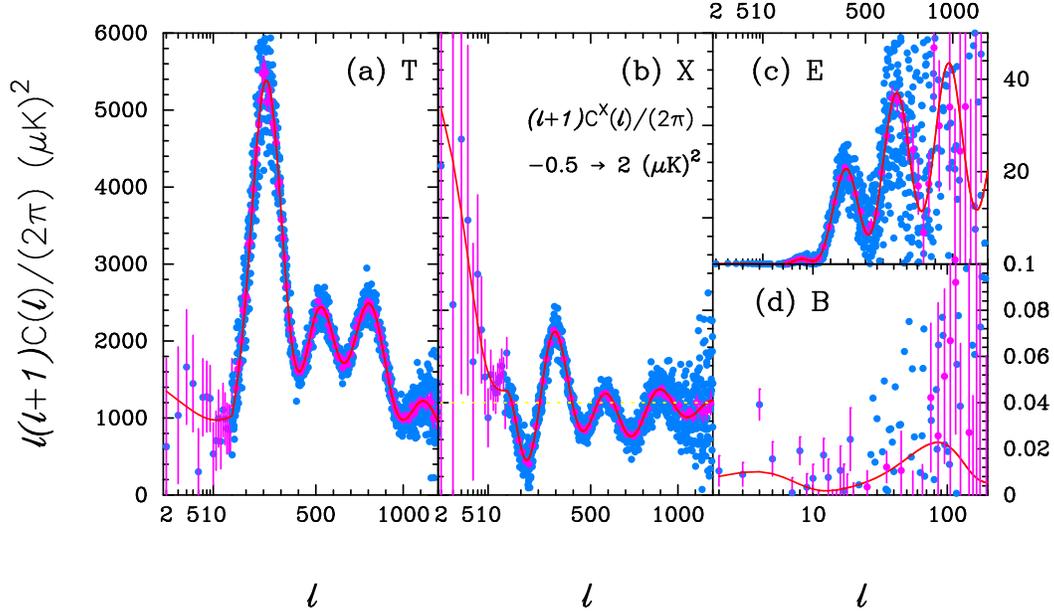}     
\caption{Power spectra for a simulation matched to the Planck 
$143$ GHz sensitivity. The blue points show individual points of the
power spectra. The magenta points at $\ell > 20$ show power spectra
averaged in bands of width $\Delta \ell \approx 20$.  The red lines
shows the theoretical power spectra used to generate this realisation
based on the `minimal' WMAP $\Lambda$CDM parameters
\cite{ref:spergel}, but also including a tensor mode with $r=0.15$.}

\end{figure}

\smallskip 
\noindent
{\bf [B] Polarized foregrounds will be dominant:}

\smallskip 
In the case of temperature anisotropies, WMAP and other experiments
have shown conclusively that at frequencies of around $100$ GHz
Galactic emission over much of the sky is small compared to the
primordial anisotropy signal. It is therefore possible to estimate the
temperature power spectrum accurately by masking out a relatively
small area of the sky around the Galactic plane.
The situation in polarization is very different, particularly for
large angle $B$-mode anisotropies. Outside the P06 polarization mask
defined by the WMAP team, the Galactic polarized signal at $65$ GHz
contributes $\sim 0.7 \mu {\rm K}$\footnote{To $[\ell(\ell+1)
C_\ell/(2\pi)]^{1/2}.$} at low multipoles \cite{ref:page}, compared to
an expected $B$-mode signal of $\sim 0.05 \mu K$ if $r \sim
0.1$. Thus, one must be able to subtract polarized foregrounds in $Q$
and $U$ maps to a precision of order a percent to detect a primordial
$B$-mode with $r \sim 0.1$ at low multipoles.  Foreground removal is
not a well posed problem and it is not easy to assess the accuracy of
any given technique. Even for temperature anisotropies, no experiment
has yet been done with enough frequency channels to fit and subtract a
realistic parametric model of the foreground components. Luckily,
foregrounds are unimportant for the temperature anisotropies over much
of the sky and so it doesn't matter critically how
foreground cleaning is handled.  But it certainly does matter for
polarization. The ultimate limit of polarized foreground cleaning is
not yet known and this should be borne in mind when assessing
experiments with instrumental sensitivities capable of achieving $r \sim
0.01$ or better. The dominance of polarized foregrounds at low
multipoles suggests that an effective strategy for detecting $B$-modes
is to target the multipole range $\ell \sim 50 - 100$ at frequencies
of $\sim 100\; {\rm GHz}$ in the cleanest regions of the sky. This
approach requires much higher sensitivity than Planck, but lowers the
foreground contribution compared to any primordial signal and has the
added advantage that such an experiment can be done from the
ground. This is the strategy adopted by  BICEP, QUIET and CLOVER.

\smallskip 

\noindent
{\bf [C] We do not know what to expect theoretically:}

\smallskip 

Well before the temperature anisotropies were discovered (and well
before  inflation was introduced into cosmology) theorists had
made well-motivated predictions of their amplitude and spectral shape
({\it e.g.}  \cite{ref:peebles}, \cite{ref:sunyaev}). Furthermore,
under the general paradigm of an inflationary $\Lambda$CDM cosmology
it is possible to compute the $TE$ and $EE$ power spectra accurately.
We therefore have a good idea of what to expect for the scalar mode
contribution to the polarization of the CMB,  ahead of any experiment.
But we have no such guidance for the $B$-mode anisotropies from tensor
modes.

Consider for, for example, the simple power-law potential
\footnote{We use natural units, $c= \hbar =
1$. The reduced Planck mass is $M_{pl} = (8 \pi G)^{-1/2} = 2.44
\times 10^{18} {\rm GeV}$ and will be set to unity unless explicitly stated
otherwise.}
\begin{equation}
V(\phi) =   \lambda \phi^\alpha, \label{theory2}
\end{equation}
advocated by Linde \cite{ref:linde} and others.  Inflation occurs for
field values, $\phi \simgt \alpha/ \sqrt 2$, and the observed
amplitude of the scalar fluctuations in our Universe can be reproduced
for suitably small values of the parameter $\lambda$.  (For example,
the quartic potential requires $\lambda \sim 4 \times 10^{-14}$).  In
this model, the tensor-scalar ratio and scalar spectral index are given
by,
\begin{equation}
r \approx  {4 \alpha \over N} \approx {\alpha \over 15}, \quad n_s \approx 1 - {2 + \alpha \over 2 N}
\approx 1 - {2 + \alpha \over 120}, \label{theory3}
\end{equation}
where $N$ (assumed to be $\approx 60$ \cite{ref:liddle}) is the number
of inflationary e-foldings between the time that CMB scales `crossed'
the Hubble radius and the end of inflation.  The quartic potential is
marginally excluded ($\sim 3 \sigma$) by the WMAP constraints on $n_s$
and $r$, but a quadratic potential provides an acceptable fit
\cite{ref:spergel}. If the quadratic potential is correct, we expect
to find a significant spectral tilt $n_s \sim 0.97$ and a high
tensor-scalar ratio $r \sim 0.13$.

The power-law models are particular examples of so called `high-field'
inflation models (since the inflaton changes by $\Delta \phi \sim
(N\alpha)^{1/2}$ in Planck units during the last $N$ e-folds of
inflation). In fact, to produce a detectable tensor component in any
foreseeable experiment, inflation must necessarily
involve large field variations $\Delta \phi \simgt 1$ (sometimes called the
`Lyth bound' \cite{ref:lyth}, \cite{ref:emack}).

Steinhardt and collaborators \cite{ref:stein1} \cite{ref:stein2}
have argued that high tensor amplitudes must be expected unless the
inflationary potential is unnaturally finely tuned. The argument goes
roughly as follows: fluctuations on
CMB scales were frozen $N \sim 60$ e-folds from the end of inflation
when the field was rolling slowly,  $\epsilon = 2(V^\prime/V)^2 \ll 1$.
Now assume that inflation ends when slow-roll is
violated ($\epsilon =   1$).  This sets `natural'
values for the gradients of the inflaton potential,
\begin{equation}
 {V^\prime \over V} \sim {V^{\prime\prime} \over V^\prime} \sim {1
 \over N}, \label{theory4}
\end{equation}
leading to the expectations,
\begin{equation}
r \sim  {14 \over N}, \quad n_s \sim 1 - {3 \over  N}, \label{theory5}
\end{equation}
similar to the predictions of the power law model (\ref{theory3}).

How seriously should we take this argument? My view is that we should
be skeptical of such simplistic arguments in the absence of a
fundamental theory of inflation.  For example, in the last few years
there has been a lot of interest in the idea of brane inflation (see
{\it e.g.} \cite{ref:tye} for a review). In one particular scenario,
inflation arises as a D3 brane moves towards an anti-D3 brane (or
stack of anti-branes) sitting at the bottom of a 
`throat' in the flux compactified bulk in Type IIB string theory
\cite{ref:kachru}. There are many variants of this idea: the bulk may
have many throats (with the standard model brane lying in a different
throat to the `inflationary' branes); inflation may occur by the usual
slow-roll mechanism, or via the `UV' Dirac-Born-Infeld (DBI) mechanism,
in which the velocity of the inflation branes is fixed by the limiting
speed in the warped geometry.  It may be possible to realise `IR' DBI
inflation in which branes roll out of a throat into the bulk.  The
phenomenology of these types of models has been discussed recently in
\cite{ref:bean1}, \cite{ref:bean2}, to which we refer the interested
reader. These models must satisfy a geometrical constraint on the
volume of the warped throat \cite{ref:baumann} which,  in turn,  sets a
limit on the inflaton field variation during inflation of
\begin{equation}
\Delta \phi  <  \left ({4 \over N_C} \right )^{1/2}.  \label{theory6}
\end{equation}
Here,  $N_C$ is the background number of charges which is usually $N_c
\gg 1$ in this type of model. Applying the Lyth bound, the constraint
(\ref{theory6}) restricts the tensor-scalar ratio to be small. In
\cite{ref:bean1} some specific parameter combinations were found which
 gave $r$ as high as $r \sim 4 \times 10^{-3}$, but generically $r$
is expected to be many orders of magnitude smaller and hence
unobservable. For the `IR' DBI models, the analysis of
\cite{ref:bean2} suggests $r \simlt 10^{-13}$.  Kallosh and Linde
\cite{ref:kallosh} have argued for an even more stringent bound of $r
\simlt 10^{-24}$ on string inflation models by imposing the
requirement that the Hubble constant during inflation is smaller that
the gravitino mass, which they assume is in the TeV range. Possible
ways of evading the geometrical and gravitino bounds are discussed in
\cite{ref:bean1} and \cite{ref:kallosh}, but the key point 
is that at present there are no compelling theoretical predictions for
the amplitude of the tensor component. Phenomenological models can be
constructed which lead to a high tensor amplitude, while brane
inflation models can be constructed with 
unobservably small tensor amplitudes.

Given this theoretical uncertainty, is it worthwhile building
experiments that cover only a small range in $r$?  My
view on this is as follows (see \cite{ref:efstathiou2} for a more
detailed discussion). It is feasible to design experiments (at
relatively low cost) to probe a tensor-scalar ratio as low as $r \sim
10^{-2}$. By probing the multipole range $\ell \sim 50 - 100$, it
should be possible to monitor the accuracy of foreground removal by
demonstrating the reproducibility of any putative detection in
different clean regions of the sky,  and as a function of frequency.
A failure to detect tensor modes at this level would rule out 
`chaotic' inflationary models such as (\ref{theory2}) and other 
examples of `high-field' inflation. This is a well-motivated and
achievable goal.

But if we fail to detect tensor modes at the level of $10^{-2}$, what
then?  Do we continue the search with more complicated experiments
(perhaps a new CMB satellite). An experiment to probe $r \sim 10^{-4}$
would be formidably difficult, yet such an experiment would improve
the constraints on the energy scale of inflation by a small factor of
only $\sim 3$ (equation \ref{theory1}). In my view, the case for such
an experiment is weak unless there are strong theoretical reasons to
favour this narrow energy range. 

So, my strategy would be to pursue an aggressive experimental campaign
to achieve a limit $r \sim 10^{-2}$, since this seems feasible and
would rule out an important class of inflationary models. But without
strong theoretical motivation, I would not blindly continue the search
to still lower amplitudes.  In this case, it may well be more
profitable to test models of inflation by designing experiments to
detect, for example, non-Gaussianity or signatures of cosmic strings.

In the last few years we have become used to a wealth of new high
precision information from CMB experiments. The temperature
anisotropies have been relatively easy to analyse and they have  had
important consequences for many areas of astrophysics and cosmology,
{\it e.g.}  theories of the early Universe, models of structure
formation, primordial nucleosynthesis, neutrino masses, ages of the
oldest stars and so on. Experiments designed to probe tensor modes are
very different, as summarized in the following table. They are
technically demanding, will be difficult to analyse, and may well lead
to a null result which, though of interest to inflationary model
builders, would likely have little impact on the wider astrophysics
community.

\begin{center}
  \begin{tabular}{c|c} 
\hline Temperature Anisotropies & Large-Scale B-mode Anisotropies  \\ 
\hline Broad science case: early and
    late universe & Narrow science case: test of inflation \\ Know
    what to aim for & No idea of    what to expect \\
High signal &  Small signal \\ \
Foregrounds unimportant & Foregrounds dominant \\ \hline
\end{tabular}
\end{center}

\section{Dynamical Dark Energy or Cosmological Constant?}

The discovery of an accelerating Universe \cite{ref:riess},
\cite{ref:perlmutter} has led to an explosion of papers on the
phenomenology of `dark energy' (for a recent review see {\it e.g.}
\cite{ref:copeland}). On the observational side a large number of
ambitious projects have been proposed to constrain the equation of
state of dark energy and its possible evolution (summarized concisely
in the {\it Report of the Dark Energy Task Force},
\cite{ref:albrecht}). However, phenomenology should not be confused
with fundamental physics. The fact that it is easy to construct a
bewildering variety of models dynamical dark energy does not mean that
any of them will turn out to be right. In my view, much of the
astronomical community has got the problem of dark energy out of
perspective. There is no sound theoretical basis for dynamical dark energy,
whereas we are beginning to see an explanation for a small
cosmological constant emerging from string theory. Furthermore,
observational data {\it favour} a cosmological constant over dynamical
dark energy.  In this Section I present three arguments for why the
cosmological constant should be given higher weight as a candidate for
the dark energy than phenomenological dynamical models.

\subsection{Occam's Razor}

For a scalar field $\phi$ rolling in a potential $V(\phi)$, the condition that 
dark energy begins to dominate at the present time 
requires the well known fine-tuning,
\begin{equation}
  V(\phi_0)  \sim 3 H_0^2 \sim 1.2 \times 10^{-120} \sim (10^{-3} {\rm eV})^4.  \label{T1}
\end{equation}
This fine-tuning is imposed by fiat in all models of dynamical 
dark energy. For generic potentials, if the scalar field is to show
interesting dynamical behaviour, it must be nearly massless \cite{ref:carroll}
\begin{equation}
  m_\phi = \left ( {V^{\prime\prime} \over 2} \right )^{1/2} \simlt  H_0 
 \sim (10^{-33} {\rm eV}).  \label{T2}
\end{equation}

Now, the observational constraints on the equation of state parameter
$w = p/\rho$ are already closing in on the cosmological constant value
 $w = -1$. For example, Spergel {\it et al.} \cite{ref:spergel}
combine WMAP observations of the cosmic microwave anisotropies with
the Supernova Legacy Survey \cite{ref:astier} and find $w =
-0.967^{+0.073}_{-0.072}$. Similar limits on $w$ have been found by
other authors using a variety of cosmological data sets and
theoretical assumptions ({\it e.g.} \cite{ref:liddle1} \cite{ref:wood}
\cite{ref:wright} \cite{ref:wang}). If the dark energy is a scalar field,
then the field must be moving slowly
${1 \over 2} \dot \phi^2/V \ll 1$. The equation of motion
of the scalar field then imposes a constraint on the derivative
of the potential,
\begin{equation}
\left \vert { V^\prime \over
  V} \right \vert \approx \sqrt{3} \left (1 + w_{\phi_0} \over \Omega_{\phi_0} \right
)^{1/2}.  \label{T3}
\end{equation}
A value of $w_{\phi_0}$ close to $-1$ therefore requires yet another
fine-tuning in addition to those of (\ref{T1}) and (\ref{T2}), namely
that the derivative of the potential is small in Planck
units. Consider, for example, the archetypal `tracker' potential
\cite{ref:steinhardt1}
\begin{equation}
  V(\phi) = M^{4+\alpha} \phi^{-\alpha}.  \label{T4}
\end{equation}
The attractor solutions for this potential, subject to the constraint
$\Omega_{\phi_0}=0.72$, give $w_{\phi_0} =-0.411$, $-0.495$, $-0.643$,
$-0.768$, $-0.864$ for $\alpha = 6$, $4$, $2$, $1$ and $0.5$. If the
experimental constraints continue to tighten around $w = -1$, the fine
tuning required by (\ref{T3}) becomes even more acute. For example, if we
could constraint $w$ to lie in the range $-1 \simlt w \simlt -0.97$, the attractor solutions
of (\ref{T4}) would require $\alpha < 0.1$, which most readers must
surely find contrived.

Occam's razor suggests that a cosmological constant is a more
economical explanation of the observational data than a dynamical
model carefully chosen to satisfy the fine tunings of (\ref{T1})-
(\ref{T3}). Some authors have attempted to quantify Occam's razor by
computing Bayesian Evidence or related information criteria (for example
 \cite{ref:szydlowski}, \cite{ref:liddle1}, \cite{ref:sahlen},
\cite{ref:serra}, \cite{ref:davis}). However, in the cosmological
context Bayesian Evidence is difficult to compute and sensitive to
assumptions concerning prior distributions of parameters \footnote{For
  inference problems in which Bayesian Evidence is more easily
  interpreted see \cite{ref:mackay}.} It is therefore difficult to get a
precise measure of how much the data favour model `A' over model `B', but
evidently the more the data forces us towards $w=-1$, the better the
case for a cosmological constant compared to more complex models.

It is also worth noting that if, after a lot of hard work, the
observations tighten around $w =-1$ to high precision (say to an
accuracy of a percent or so), we will not be able to rule out
dynamical dark energy. Such a constraint at low redshift ($z \sim
0.5$, depending on the nature of the cosmological data, {\it e.g.}
supernovae, lensing, baryon oscillations) will constrain the potential
to be very nearly flat for field values $\phi \sim \phi_0$ (equation
\ref{T3}), but it is easy to construct dynamical models in which the
potential changes shape outside the redshift window probed by the data
leading to an abrupt change in $w_\phi$ \cite{ref:chongchitnan}.  A
double exponential potential,
\begin{equation}
  V(\phi) = M_1^4 {\rm e}^{-\lambda_1 \phi} + 
M_2^4 {\rm e}^{-\lambda_2 \phi}, \label{T5}
\end{equation}
with $\lambda_1 \gg \lambda_2$, $\lambda_2 \ll 1$ provides an example
\cite{ref:barreiro}. With a double exponential it is possible to
construct models which make a transition from the attractor solution
$\Omega_\phi = 3(1 + w_B)/\lambda_1^2$, $w_\phi = w_B$, at high
redshift (where $w_B$ is equation of state of the background fluid) to
the late time attractor at low redshift with $w_\phi = - 1 +
\lambda_2^2/3$, $\Omega_\phi = 1$. The parameters of this type of
model can easily be adjusted to give a dynamically significant dark
energy density at high redshift while mimicking a cosmological
constant to arbitrary precision at low redshift. No matter how precise
the observations become, we will always be able to construct models of
this sort \cite{ref:chongchitnan}. But, of course, without strong
theoretical motivation, such models are not likely to be taken
seriously. I mention this point here to emphasize that one cannot take
a purely empirical approach to the problem of dark energy. Dark energy
surveys must be assessed within a theoretical (and not purely 
phenomenological) framework.

\subsection{The String Landscape}

A fundamental theory of the cosmological constant must involve quantum
gravity. Quantum mechanics is required to endow the vacuum with an
energy density and gravity is required if we are to `feel' the effects
of this vacuum.  At present, string theory is our best bet for a
consistent quantum theory of gravity, so it is reasonable to ask
what string theory has to say about dark energy.  There has been
substantial progress on this question in the last few years (see, for
example, the reviews by Polchinski \cite{ref:polchinski} and Bousso
\cite{ref:bousso1}, \cite{ref:bousso2}). As is well known, string
theory requires either $9+1$ or $10+1$ spacetime dimensions.  Six or
seven spatial dimensions must therefore be compactified so that they remain
hidden from us.  It now seems that a huge number, perhaps $10^{500}$
or more, metastable vacua may exist depending on the choice of compact
manifold and different values of magnetic fluxes wrapped over
different homology cycles \cite{ref:bousso3} \cite{ref:kachru1}. 
These vacuum solutions form the so-called `landscape' of string
theory \cite{ref:susskind}.

The existence of a landscape of vacua raises the possibility of an
{\it anthropic} explanation for the cosmological constant, as first
advocated by Weinberg \cite{ref:weinberg}.  To some people, anthropic
reasoning is abhorrent and represents a retreat from conventional
science. I disagree with this view. An anthropic explanation for
$\Lambda$ requires a very special theoretical framework, placing
restrictive conditions on fundamental theory. In particular
\cite{ref:bousso2}:

\begin{itemize}

\item
the value of $\Lambda$ must `scan', either
continuously or with sufficiently close spacing to account for the
small value of $10^{-120}$;

\item
vacua with small values of $\Lambda$ must be realised  
({\it i.e.} populated);

\item
vacua must exist that are consistent with the physics of the standard model;

\item
they should admit a period of inflation to produce a big
Universe containing matter, radiation and the fluctuations necessary
to form non-linear structure by the present day.

\end{itemize}

These are non-trivial conditions.  Although we do not yet know the
details of how these conditions might be satisfied\footnote{In
  addition, there is still no compelling solution of the `measure
  problem' that bedevils the interpretation of eternal inflation and
  the string landscape (see {\it e.g.} \cite{ref:garriga},
  \cite{ref:vilenkin1}, \cite{ref:linde1}).}, the landscape of string
vacua at least offers the {\it possibility} of a theoretical framework
within which they may be met. String theory was not designed to solve
the cosmological constant problem, yet it may contain all of the
ingredients necessary to realize Weinberg's anthropic prediction of
$\Lambda$. If this is correct,  $\Lambda$ must be much more
strongly favoured than dynamical dark energy.

\subsection{Back to square one} The smallness of the cosmological
constant is a fundamental problem in theoretical physics. We must
therefore understand what we are doing when we construct
phenomenological models of dark energy. Why should we write down a
potential such as (\ref{T4}) rather than the potential
\begin{equation}
  V(\phi) = M^{4+\alpha} \phi^{-\alpha} + V_0?  \label{S1}
\end{equation}
In other words, why should we assume that the cosmological constant is
zero in writing down a potential for dark energy? Despite many years
of effort no mechanism has been found that can enforce $V_0$ to be
zero \cite{ref:weinberg2}. In fact there are good theoretical reasons
to suggest that no such mechanism exists \cite{ref:bousso2}. The
assumption that $V_0$ is zero is therefore not benign. It raises the
fundamental question of why the vacuum energy and `bare' cosmological
constant cancel exactly, to which there is no known answer.


\section{The Non-Linear Universe}

We have learnt a lot about fundamental physics by studying the
non-linear Universe at low redshifts. Highlights include the discovery
of dark matter in galaxies and clusters, evidence for the hierarchical
assembly of galaxies (as expected in the cold dark matter model) and
the discovery of the accelerating Universe. Will this link between
fundamental physics and astronomy remain as strong in the future?
Perhaps not.  Unless there are some new surprises, we probably already
know enough about the initial fluctuations and cosmological parameters
to predict the low redshift Universe. Understanding the formation of
the first stars, or the formation and evolution of galaxies, then
becomes an exercise in complex non-linear physics, of no more
relevance to fundamental physics than weather forecasting.

On the other hand, there is plenty of scope for surprises,  for example:
\begin{itemize}
\item evidence of topological defects, such as cosmic superstrings
  \cite{ref:polchinski2};
\item clues to the nature of the dark matter {\it e.g.} evidence of 
dark matter annihilation or direct laboratory detection;
\item
firm evidence for dynamical dark energy;
\item
evidence of non-minimal coupling between dark matter and dark energy;
\item
non-Gaussianities in the primordial fluctuations, perhaps indicative
of brane inflation \cite{ref:silverstein};
\item 
features, such as a large spectral index variation or a sharp
  spike, in the fluctuation spectrum;
\item
evidence of modifications to General Relativity, perhaps associated
with higher dimensional physics \cite{ref:dvali};
\item
observational signatures of other universes \cite{ref:aguirre}.
\end{itemize}

Any one of the above would constitute an important discovery, and some
might well be considered revolutionary. But there is no guarantee that
we will discover any of these things.  Suppose that several years from
now the WMAP `concordance' cosmology still holds up, what then?  The
non-linear Universe may then be of little interest to fundamental
physicists, but it nevertheless poses problems that are interesting
{\it in their own right}. Finding extra-solar planets, understanding
how they form and whether they harbour life are interesting problems,
though they will not tell us anything new about fundamental
physics. Similarly it is important to develop an understanding of the
rich phenomena that we observer in the Universe such as, supernovae,
supermassive black holes, quasars, galaxies, gamma-ray bursts.
Astronomers should offer no apologies about studying Nature in all her
complex glory.

\section{Prognosis}

\subsection{The Future of Precision Cosmology} We have seen very
remarkable progress in the last few years in constraining theoretical
models and determining cosmological parameters using a diverse set of
astronomical surveys. There have been some problems along the way, for
example, the apparent high optical depth for reionization in the first
year WMAP data \cite{ref:bennett} and tension between different
supernovae and galaxy redshift data sets \cite{ref:nesseris},
\cite{ref:percival}, but these have not been too serious. We have been
lucky. The next generation of large surveys will be aimed at much more
difficult problems, such as constraining $w$, absolute neutrino
masses, $B$-mode anisotropies {\it etc} and there is no guarantee that
systematic errors can be controlled to the required level of
precision. Progress is therefore likely to be slower than we have
become used to, with more false results. I would advocate the
following:

\noindent
(i) new surveys should be designed with as many internal
 redundancy checks as possible; 

\noindent
(ii) perform complementary surveys since consistency between different
types of astrophysical data provides powerful tests of systematics;

\noindent
(iii) when making the case for ambitious new experiments do not
succumb to \"uber-forecasting -- include realistic errors,
foregrounds, systematics {\it etc}.

\subsection{The Future of CMB Experiments}
Post Planck, most CMB experiments are targeting either high
resolution observations of secondary anisotropies (the
Sunyaev-Zeldovich effect in particular) or low resolution observations
of B-mode anisotropies.  However, the science case for $B$-mode
experiments is highly specialised and so we need to think carefully
about their likely impact. As discussed in Section 2, sub-orbital and
ground based experiments aimed at achieving tensor-scalar ratios of
$\sim 10^{-2}$ are well motivated. Detection of tensor modes above
this level would constitute a major discovery, providing firm evidence
that inflation took place and fixing the energy scale of
inflation. Failure to detect tensor modes at the $\sim 10^{-2}$ level
would rule out an important class of models that has played an
influential role in the development of inflationary cosmology. But we
need to think carefully before pressing the case for a new CMB
satellite to detect a tensor-scalar ratio in the range $10^{-2} \simgt
r \simgt 10^{-4}$. Bearing in mind that a low resolution polarization
satellite is a single goal mission, that foreground polarization
is dominant, and systematics need to be controlled to an unprecedented
level, it only makes sense to target this narrow range in $r$
{\it if there are compelling theoretical reasons to do so}. Otherwise,
a failure to detect a tensor mode would have little scientific
impact. The science cases for other types of CMB experiments are
difficult to judge at this stage. Hints of non-Gaussianity, or cosmic
strings, could motivate a new generation of CMB experiments with high
potential for scientific discovery.

\subsection{Unravelling The Nature of Dark Energy} In Section 3, I
presented arguments for why the cosmological constant should be
strongly favoured over more exotic dark energy candidates such as
quintessence. Does this mean that there is no point in supporting new
dark energy surveys \cite{ref:albrecht1}? Of course not! The arguments
presented in Section 3 are not intended to dissuade people from
ambitious programmes to probe the nature of dark energy. But they are
intended to influence the nature of the programmes themselves. If you
are at all persuaded by the arguments of Section 3, then you should
expect that future experiments will simply strengthen the case for a
cosmological constant. By all means design experiments to test for
dynamical dark energy, but expect failure!
Dark energy
surveys should therefore be designed to have a broad {\it
  astrophysical} science case, so that if we find nothing
fundamentally new about dark energy we will at least learn something
interesting about astrophysics. Simon White, using quite different
arguments based on the sociological implications of dark energy
surveys,  has reached similar conclusions. Surveys narrowly focused on
dark energy will have little scientific impact on both astronomy and
theoretical physics if they merely tighten the limits around
$w=-1$. They will have even less impact if systematic errors prevent
them from achieving their stated goals.

\subsection{The Non-Linear Universe } We are extremely fortunate in
having a large array of expensive facilities with which to study the
Universe. We are even more fortunate to have major new telescopes and
observatories on the horizon, including ALMA, Pan-STARRS, Herschel,
JWST, LSST, LISA, ELTs and so on. This diverse range of facilities
guarantees that astronomy will remain a vibrant subject for many years
to come. In Section 4, I listed ways in which new physics could
influence what we see today. Most of these effects are so speculative
that the likelihood of observing any one of them is very small.  In
judging scientific projects, one must strike a balance between the
importance of a scientific discovery against the likelihood of making
the discovery in the first place.  In my view, we should continue
using our generous array of facilities, together with theoretical
insight and increasingly powerful computers, to build up a picture of
how complex objects in the Universe formed and evolved. Astronomy
should not succumb to fundamentalism -- understanding the complexity
of our Universe is an important problem in its own right. We therefore
need to maintain a diverse range of research programmes, rather than
assigning resources to projects with narrow and highly speculative
science goals. If we maintain diversity, surprises of relevance to
fundamental physics will surely come.

The story of astronomy is one of unexpected discovery after unexpected
discovery. This is why our subject is so interesting. It is extremely
unlikely that the `concordance' $\Lambda$CDM model is the last word in
cosmology.  There will be surprises in store and they will have
revolutionary implications for fundamental physics. Is my confidence
in this an example of irrational exuberance? Perhaps, but history is
definitely on my side.


\end{document}